# Comments on the paper: An extensive investigation on nucleation, growth parameters, crystalline perfection, spectroscopy, thermal, optical, microhardness, dielectric and SHG studies on potential NLO crystal – Ammonium Hydrogen L-tartarte [Spectrochim Acta 103A (2013) 388-399]

Bikshandarkoil R. Srinivasan, V.S. Nadkarni
Department of Chemistry, Goa University, Goa 403206, INDIA
Email: srini@unigoa.ac.in (BRS) nitin@unigoa.ac.in (VSN)

**Abstract**

The authors of the title paper report to have grown ammonium hydrogen *L*-tartrate (**1**) by use of *L*-glutamine and *L*-tartaric acid in 1:1 mole ratio. Many points of criticism pertaining to the characterization of **1** are described herein.

**Introduction**

Tartaric acid and its salts have played a key role in the discovery of chemical chirality. The research history and the contribution of tartaric acid and its compounds to science has been reported by Derewenda [1]. Tartaric acid is a dihydroxy-dicarboxylic acid and can exist as *L*-(+)-tartaric acid (natural isomer), *D*-(–)-tartaric acid (unnatural isomer), mesotartaric acid, *DL*(±)-tartaric acid. The *L*-(+) and *D*-(–) forms of tartaric acid are chiral (enantiomers) and can be distinguished by their optical rotation, while the achiral mesoform is a diastereoisomer of both the chiral isomers. It is well documented that the diammonium salt of *L*-tartaric acid can be synthesized by reaction of an alcoholic solution of the diacid with gaseous ammonia [2]. Ammonium hydrogen *L*(+)-tartrate (**1**) also known as ammonium (+) bitartrate is synthesized by the reaction of the diammonium salt with *L*-tartaric acid [3] and its optical rotation has been reported [4]. A three component crystal based on a central ammonium (+) bitartrate first described by Louis Pasteur has been recently structurally characterized by Wheeler et al [5]. However the title paper does not add any new information on compound **1** as shown below.

**Comment**

Although, ammonium hydrogen *L*(+)-tartrate (**1**) is a commercially available solid [6], the authors of the title paper [7] claim to have performed an aqueous reaction of *L*-tartaric acid with *L*-glutamine in 1:1 mole ratio to prepare (**1**) in order to grow big crystals. In view of the fact that the authors described in their introduction about amino acids as interesting materials for NLO applications, it appears that they used *L*-glutamine (an amino acid) as a source of ammonia for synthesis of **1**. The reason for choosing *L*-glutamine which is a costly source for ammonia is unclear. Although *L*-glutamine which is an amide is known to get hydrolysed in acidic medium with the liberation of ammonia, the other product of the reaction is *L*-glutamic acid. This aspect has not been taken into account by the authors since no mention is made of either *L*-glutamic acid or the steps taken to avoid its contamination of **1**.

Our above argument namely an impure product is confirmed by the reported $^1$H NMR spectrum in the title paper. The spurious signals and the values of integration for H atoms indicate that the composition of the reported compound is not that of **1** but indeterminate. A comparison of the reported spectrum with the expected $^1$H NMR spectrum of **1** predicted by using a spectrum predictor reveals the dubious nature of the grown crystal. In addition to not being aware of the normal practice of assigning signals of O-H protons based on D$_2$O exchange experiment, the authors have made unusual assignments for example assigning the quartet at ~ δ 3.96 for (NH$_4$)$^+$ proton and a signal at ~ δ 1.24 ppm in the reported spectrum for C-C bonding of *L*-tartaric acid. Such assignments of $^1$H NMR data are totally unheard of and reveal that the authors do not have the necessary expertise to interpret NMR spectral data.

The dubious nature of the grown crystals can also be evidenced from the elemental data of C, 42.52, N, 8.31 and O, 49.17% in the title paper under the title SEM-EDX study. The analytical values for C, N and O add up to 100% indicating that the compound under investigation does not contain any hydrogen. In spite of this, the authors reported that they have obtained qualitative and quantitative determination of the elements present in the sample. Here again the authors are unaware that EDX is an inappropriate method for analysis of lighter elements like C, H, N etc.

We find the claim of the authors that the unit cell parameters of the grown crystal are in good agreement with that of the reported [8] unit cell of ammonium hydrogen *L*(+)-tartrate quite surprising in view of an inconsistent NMR spectrum for the title compound. Since no details of single crystal structure determination other than the unit cell data are given in the title paper, it is not clear if this experiment was really performed. It is also noted that the authors have assigned the Sohncke $P2_12_12_1$ space group without either giving the value of Flack parameter or giving the reasons for choosing this space group based on systematic absences. Such a practice of reporting structures of compounds based only on unit cell data has more often resulted in erroneous conclusions, a point which has been well demonstrated in the literature [9-12].

The synthesis, properties and structure of the chiral compound **1** which is commercially available [6], is well documented in the literature. Unfortunatelty the authors have not taken into account that the purity of an optically compound like **1** is to be confirmed by measuring its optical rotation independent of a structure determination. In the present case the optical purity needs to be unambiguously established in view of an optically active by product namely *L*-glutamine. However no such study was performed. In the absence of such characterization, the only option left is a determination of the crystal structure, to unambiguously establish the exact nature of the product formed from the interesting (?) reaction of L-tartaric acid (dicarboxylic acid) and L-glutamine (an amide).

In view of the questionable nature of the grown crystals of **1**, other reported studies like the microhardness, dielectric and SHG for this so called potential NLO crystal are meaningless and do not merit any discussion.